\documentclass[10pt,twocolumn,pra,nofootinbib,notitlepage]{revtex4-1}
\usepackage[latin9]{inputenc}
\usepackage{amsmath}
\usepackage{amssymb}
\usepackage{graphicx}


\begin{document}

\title{General linearized theory of quantum fluctuations around arbitrary
limit cycles}

\author{Carlos Navarrete-Benlloch$^{1,2}$, Talitha Weiss$^{1,2}$, Stefan
Walter$^{1,2}$, and Germán J. de Valcárcel$^{3}$}

\affiliation{$^{(1)}$Max-Planck-Institut für die Physik des Lichts, Staudtstrasse
2, 91058 Erlangen, Germany}

\affiliation{$^{(2)}$Institute for Theoretical Physics, Erlangen-Nürnberg Universität,
Staudtstrasse 7, 91058 Erlangen, Germany}

\affiliation{$^{(3)}$Departament d\textquoteright \`Optica, Facultat de F\'isica,
Universitat de Val\`encia, \emph{Dr}. Moliner 50, 46100 Burjassot,
Spain}

\begin{abstract}
The theory of Gaussian quantum fluctuations around classical steady
states in nonlinear quantum-optical systems (also known as standard
linearization) is a cornerstone for the analysis of such systems.
Its simplicity, together with its accuracy far from critical points
or situations where the nonlinearity reaches the strong coupling regime,
has turned it into a widespread technique, which is the first method
of choice in most works on the subject. However, such a technique
finds strong practical and conceptual complications when one tries
to apply it to situations in which the classical long-time solution
is time dependent, a most prominent example being spontaneous limit-cycle
formation. Here we introduce a linearization scheme adapted to such
situations, using the driven Van der Pol oscillator as a testbed for
the method, which allows us to compare it with full numerical simulations.
On a conceptual level, the scheme relies on the connection between
the emergence of limit cycles and the spontaneous breaking of the
symmetry under temporal translations. On the practical side, the method
keeps the simplicity and linear scaling with the size of the problem
(number of modes) characteristic of standard linearization, making
it applicable to large (many-body) systems.
\end{abstract}
\maketitle
\textbf{Introduction}. The advent of modern quantum technologies has
triggered the discovery of a plethora of optical, atomic, and solid
state devices working in the quantum regime \cite{Dowling03} (see
also the starting paragraph of \cite{Benito16}\textbf{ }and the references
therein). A first-principles approach leads to a description of such
devices as open quantum systems evolving according to nonlinear Hamiltonians
and incoherent processes like dissipation \cite{GardinerZoller,BreuerPetruccione,Carmichael1,GardinerZollerI}.
Mathematically, one has to face master equations for the state of
the system or quantum Langevin equations for its operators, which
are in general impossible to solve exactly.

On the other hand, quantum nonlinearities are very difficult to observe
in the laboratory and therefore most experiments are well described
by effective linear models. The most widespread method for obtaining
such linear models starting from nonlinear ones is the so-called \textit{standard
linearization }\cite{Drummond81,Lugiato81}, which consists in a \textit{Gaussian-state
ansatz} centered at the solution of the system's nonlinear equations
in the classical limit \cite{CNB14}. The method combines incredible
simplicity with pretty good accuracy in regions of the phase diagram
where the system shows a finite number of well-spaced classical attraction
points. However, it relies on two properties of the system's state
in the classical limit: It has to be \emph{stationary} and \emph{stable}
along all directions of phase space. The first condition precludes
its application to regions where the classical solutions are time
dependent (such as limit cycles \cite{StrogatzBook,GrimshawBook},
ubiquitous to, e.g., lasing, second-harmonic generation, or optomechanical
systems). The second condition excludes the possibility of applying
it to systems which, being invariant under continuous transformations
of some kind, have a classical solution which breaks that invariance
via \textit{spontaneous symmetry breaking}. This is because Goldstone's
theorem implies the existence of a zero eigenvalue of the linear stability
matrix, and hence a direction of phase space which is not damped \cite{CNBthesis,CNB08,CNB10,Perez06,Perez07}.

While standard linearization has been generalized to deal with spontaneous
symmetry breaking of spatial, polarization, and phase symmetries \cite{CNBthesis,CNB08,CNB10,Garcia09,Garcia10,Lane88,Perez06,Perez07,Reid88,Reid89},
an extension capable of dealing with limit cycles remains. In the
case of spontaneous symmetry breaking the trick consists on using
a phase-space representation of the state to keep track of the phase-space
variable associated to the system's invariance, which will carry the
largest part of the fluctuations. Then, the theory can be linearized
with respect to any other phase-space variable.

In this work we generalize standard linearization to regions where
the classical long-time solution is time dependent, in particular
describing a periodic orbit in phase space. Our idea relies on the
connection between the emergence of such limit cycles, and the spontaneous
breaking of a very particular symmetry: arbitrary translations in
time.

For convenience, in this work we introduce the method for single-mode
problems, using the driven quantum Van der Pol (VdP) oscillator \cite{Walter14,Weiss17,Lorch16}
as an example. The simplicity of this model will allow for comparisons
with full numerical simulations. The generalization to multi-mode
problems is straightforward, and will be explored in the future for
more practical and complex problems such as optomechanical cavities
deep into the parametric instability regime \cite{Lorch14,Qian12}.
Moreover, the complexity of the method scales only linearly with the
number of modes, providing then an efficient route towards the analysis
of many-body systems out of equilibrium such as optomechanical arrays
\cite{Arrays1,Arrays2,Arrays3,Arrays4,ArraysWeiss} in the self-sustained
oscillations regime.

\textbf{Van der Pol model. }The quantum model for a driven VdP oscillator
consists of a single bosonic mode with annihilation operator $\hat{a}$,
whose state $\hat{\rho}$ evolves according to the master equation
\cite{Walter14,Weiss17}
\begin{equation}
\frac{d\hat{\rho}}{d\tau}=\left[\frac{F}{\sqrt{\gamma}}(\hat{a}^{\dagger}-\hat{a})+\mathrm{i}\Delta\hat{a}^{\dagger}\hat{a},\hat{\rho}\right]+\frac{\gamma}{2}\mathcal{D}_{a^{2}}[\hat{\rho}]+\mathcal{D}_{a^{\dagger}}[\hat{\rho}],\label{DPOmaster}
\end{equation}
where $\mathcal{D}_{J}[\hat{\rho}]=2\hat{J}\hat{\rho}\hat{J}^{\dagger}-\hat{J}^{\dagger}\hat{J}\hat{\rho}-\hat{\rho}\hat{J}^{\dagger}\hat{J}$
and the bosonic operators satisfy canonical commutation relations
$[\hat{a},\hat{a}]=0$ and $[\hat{a},\hat{a}^{\dagger}]=1$. The Hamiltonian
includes a coherent drive at rate $F/\sqrt{\gamma}>0$ detuned by
$\Delta$ with respect to the natural oscillation frequency of the
oscillator (note that we work in a picture rotating at the driving
frequency). The model contains two incoherent terms as well, the first
one corresponding to pairs of excitations lost irreversibly at rate
$\gamma$ (nonlinear losses), and the second one to linear pumping.
The rate of the latter is used to normalize the rest of rates and
frequencies, while its inverse normalizes time, so that $\tau$, $\gamma$,
$F$, and $\Delta$ are dimensionless. We show later that with these
choices the classical phase diagram of the system is determined uniquely
by $F$ and $\Delta$, while $\gamma$ determines the strength of
the quantum fluctuations.

The method is best introduced by mapping the master equation to a
set of stochastic equations. This can be done with the help of phase-space
quasiprobability distributions \cite{GardinerZoller,GardinerZollerI,Carmichael1,SchleichBook}
such as standard Wigner, Husimi, or Glauber-Sudharsan representations.
Here we choose the positive \textit{P} representation \cite{Drummond80,CarmichaelBook2,GardinerZoller,GardinerZollerI}
because, unlike the previous representations, it always leads to stochastic
equations equivalent to the master equation without any approximation.
This representation associates two independent stochastic variables
that we denote by $\beta/\sqrt{\gamma}$ and $\beta^{+}/\sqrt{\gamma}$
with the annihilation and creation operators $\hat{a}$ and $\hat{a}^{\dagger}$,
respectively, in such a way that normally-ordered quantum expectation
values and stochastic averages are related by $\langle\hat{a}^{\dagger m}\hat{a}^{n}\rangle=\langle\beta^{+m}\beta^{n}\rangle/\gamma^{(m+n)/2}$,
with $m,n\in\mathbb{N}$. Using standard techniques \cite{Drummond80,CarmichaelBook2,GardinerZoller,GardinerZollerI,GardinerBook},
we show in \cite{SupMat} that the stochastic amplitudes evolve according
to\begin{subequations}\label{LangevinEqs}
\begin{align}
\dot{\beta} & =F+(1+\mathrm{i}\Delta-\beta^{+}\beta)\beta+\sqrt{\gamma}[\sqrt{2}\xi(\tau)+\mathrm{i}\beta\eta(\tau)],\\
\dot{\beta}^{+} & \hspace{-0.6mm}=F+(1-\mathrm{i}\Delta-\beta^{+}\beta)\beta^{+}\hspace{-0.6mm}+\hspace{-0.6mm}\sqrt{\gamma}[\sqrt{2}\xi^{\ast}\hspace{-0.6mm}(\tau)\hspace{-0.6mm}-\hspace{-0.6mm}\mathrm{i}\beta^{+}\hspace{-0.6mm}\eta^{+}\hspace{-0.6mm}(\tau)],
\end{align}
\end{subequations}where $\eta(\tau)$, $\eta^{+}(\tau)$, and $\xi(\tau)$
are independent white Gaussian noises (real the first two, and complex
the last one).

\begin{figure}
\includegraphics[width=1\columnwidth]{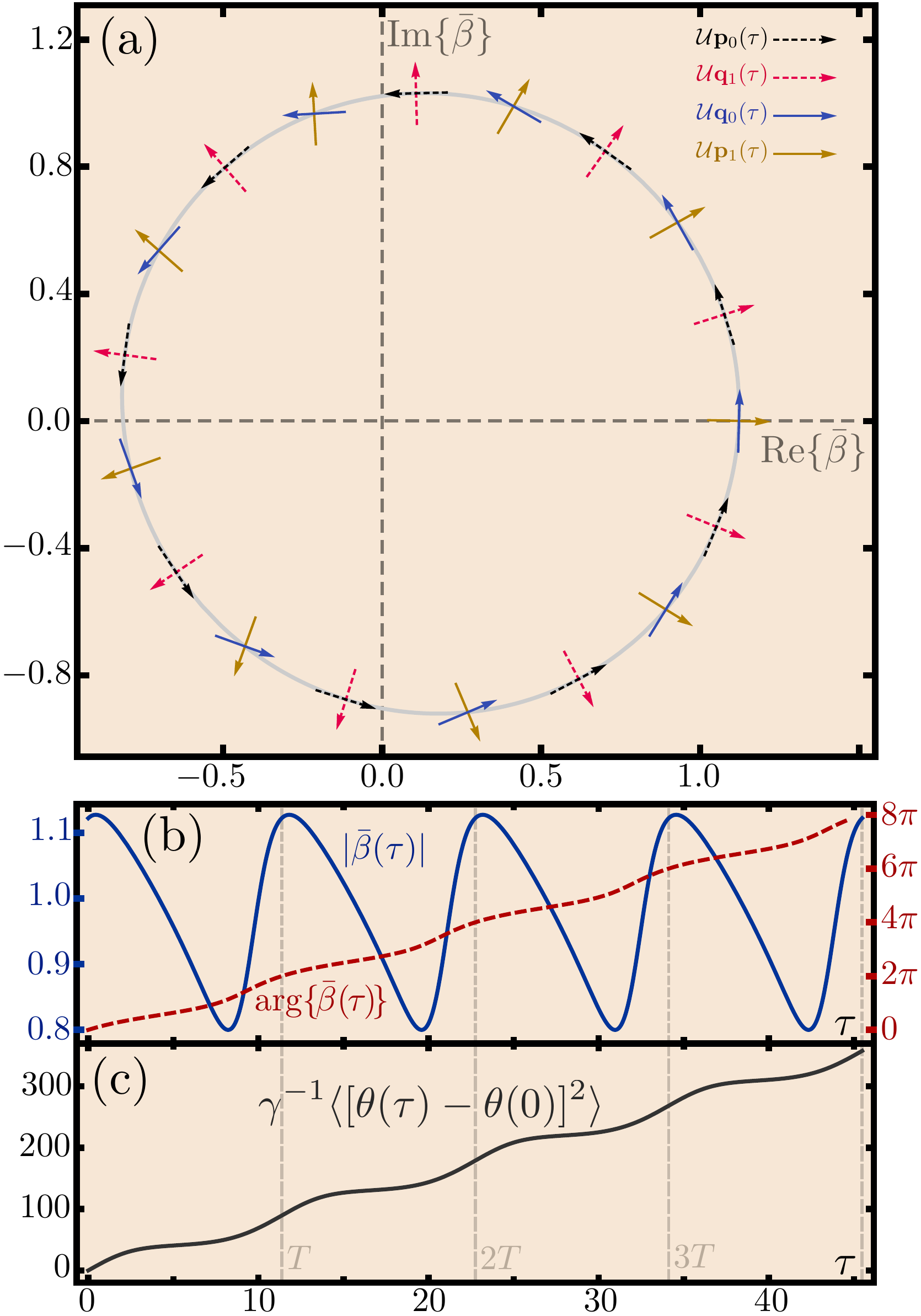}\caption{Limit cycle emerging for $\Delta=\sqrt{0.4}$ and $F=\sqrt{0.1}$.
(a) We show in grey the closed trajectory described in phase space.
The arrows refer to the direction of the Floquet eigenvectors in selected
points of the cycle. (b) Time evolution of the cycle's absolute value
and the phase. (c) Evolution of the variance of $\theta$, see Eqs.
(\ref{StochasticExpansion}) and (\ref{ThetaVariance}). Note that
$\gamma$, which sets how relevant quantum fluctuations are, appears
just as an absolute scale for the variance, whose dependence on time
is set by the limit cycle's shape. \label{Fig1}}
\end{figure}

\textbf{Limit cycles in the classical limit. }Coming from a normally
ordered representation (where vacuum noise is already taken into account
in the ordering), the equations above predict a large-amplitude coherent
state for $\gamma\rightarrow0$. We talk then about the classical
limit. The remaining deterministic equation $\dot{\beta}=F+(1+\mathrm{i}\Delta-|\beta|^{2})\beta$
is a paradigm for synchronization phenomena \cite{Weiss17}, and its
phase diagram is well known (we provide an overview of it in \cite{SupMat}).
In general terms, its stationary solutions, corresponding to solutions
oscillating at the driving frequency, are stable only provided a strong
enough drive is fed; otherwise, the oscillations are not synchronized
to the drive, so that for long times the system ends up in a nontrivial
stable periodic solution $\bar{\beta}(\tau)=\bar{\beta}(\tau+T)$
which we call \emph{limit cycle }or \emph{periodic orbit} \cite{StrogatzBook,GrimshawBook}.
In Fig.~\ref{Fig1} we show an example of such solution, where it
can be appreciated that it describes a closed curve in phase space
(a), with an absolute value and a phase that oscillate periodically
(b). Note that analytical solutions for these limit cycles exist only
in limited cases, and therefore one needs to find them numerically
in general.

\textbf{Linearization around limit cycles.} We are now able to introduce
the linearization technique for quantum fluctuations around limit
cycles. We start by expanding the stochastic amplitudes as\begin{subequations}\label{StochasticExpansion}
\begin{eqnarray}
 & \beta(\tau+\theta)=\bar{\beta}(\tau+\theta)+b(\tau+\theta),\\
 & \beta^{+}(\tau+\theta)=\bar{\beta}^{\ast}(\tau+\theta)+b^{+}(\tau+\theta).
\end{eqnarray}
\end{subequations}Here, $\theta$ determines at which point of the
cycle the solution $\bar{\beta}(\tau+\theta)$ starts for $\tau=0$,
and it is precisely the parameter which is not fixed by the classical
equations of motion: $\bar{\beta}(\tau+\theta)$ is a solution of
the equations for any choice of $\theta$. Owed to this symmetry,
quantum fluctuations cannot be considered small in arbitrary points
and directions of phase space, as nothing prevents them from acting
on $\theta$ without resistance. Hence, in order for any linearized
theory of quantum fluctuations to work, $\theta$ has to be taken
as a variable itself (making it time dependent in the expansion above)
and only then the fluctuations $b$ and $b^{+}$ can be taken as small
quantities. In addition, $\dot{\theta}$ can be taken as small quantity
as well, since variations of $\theta$ are induced by quantum noise,
which is weak in the region of interest. Introducing (\ref{StochasticExpansion})
in (\ref{LangevinEqs}), to first order in the small variables (including
noise) we then get \cite{SupMat}
\begin{equation}
\mathbf{\dot{b}}(\tau)+\mathbf{p}_{0}(\tau)\dot{\theta}(\tau)=\mathcal{L}(\tau)\mathbf{b}(\tau)+\sqrt{\gamma}\mathbf{n}(\tau)\text{,}\label{LinLan}
\end{equation}
where $\mathbf{b}=(b,b^{+})^{T}$, $\mathbf{p}_{0}=(\partial_{\tau}\bar{\beta},\partial_{\tau}\bar{\beta}^{\ast})^{T}$,
$\mathbf{n}(\tau)=[\sqrt{2}\xi(\tau)+\mathrm{i}\bar{\beta}(\tau)\eta(\tau),\sqrt{2}\xi(\tau)-\mathrm{i}\bar{\beta}^{\ast}(\tau)\eta^{+}(\tau)]^{T}$,
and
\begin{equation}
\mathcal{L}(\tau)=\left(\begin{array}{cc}
1-2|\bar{\beta}(\tau)|^{2}+\mathrm{i}\Delta & -\bar{\beta}(\tau)^{2}\\
-\bar{\beta}^{*2}(\tau) & 1-2|\bar{\beta}(\tau)|^{2}-\mathrm{i}\Delta
\end{array}\right),\label{LSM}
\end{equation}
is the linear stability matrix. Note that the noise correlations can
be written in the compact form $\langle n_{j}(\tau)n_{l}(\tau^{\prime})\rangle=N_{jl}(\tau)\delta(\tau-\tau^{\prime})$,
where $N_{jl}$ are the elements of the diffusion matrix
\begin{equation}
\mathcal{N}(\tau)=\left(\begin{array}{cc}
-\bar{\beta}^{2}(\tau) & 2\\
2 & -\bar{\beta}^{\ast2}(\tau)
\end{array}\right).
\end{equation}

As we will see, the introduction of $\theta(\tau)$ as an explicit
variable will allow us to describe properly spontaneous temporal symmetry
breaking and its associated undamped phase-space direction. 

\textbf{Floquet method and eigenvectors.} The main difference of Eq.
(\ref{LinLan}) with respect to the linearized Langevin equations
found in previous linearization methods is the time periodicity of
$\textbf{p}_{0}(\tau)$ and $\mathcal{L}(\tau)$. We deal with this
by applying Floquet theory \cite{CodingtonBook,GrimshawBook} as we
explain next.

Let us define the fundamental matrix $\mathcal{R}(\tau)$, which satisfies
the initial value problem $\dot{\mathcal{R}}(\tau)=\mathcal{L}(\tau)\mathcal{R}(\tau)$
with $\mathcal{R}(0)=\mathcal{I}$, the latter being the identity
matrix. From it, we further define the matrix $\mathcal{M}$ through
$\exp(\mathcal{M}T)=\mathcal{R}(T)$, and the $T$\emph{-}periodic
matrix $\mathcal{P}(\tau)=\mathcal{R}(\tau)\exp(-\mathcal{M}\tau)$.
Given the eigensystem $\{\mathbf{v}_{j},\mathbf{w}_{j};\mu_{j}\}_{j=0,1}$
of $\mathcal{M}$, composed of right and left orthogonal ($\mathbf{w}_{j}^{\dagger}\mathbf{v}_{l}=\delta_{jl}$)
eigenvectors satisfying $\mathcal{M}\mathbf{v}_{j}=\mu_{j}\mathbf{v}_{j}$
and $\mathbf{w}_{j}^{\dagger}\mathcal{M}=\mu_{j}\mathbf{w}_{j}^{\dagger}$,
we introduce the Floquet eigenvectors $\mathbf{p}_{j}(\tau)=\mathcal{P}(\tau)\mathbf{v}_{j}$
and $\mathbf{q}_{j}^{\dagger}(\tau)=\mathbf{w}_{j}^{\dagger}\mathcal{P}^{-1}(\tau)$.
As we show along the next sections, knowledge of these vectors is
enough to derive the linearized quantum properties of the system.
To this aim, it is also convenient to point out that they satisfy
the initial value problems\begin{subequations} \label{EigenTime}
\begin{align}
\mathbf{\dot{p}}_{j}\left(\tau\right) & =\left[\mathcal{L}\left(\tau\right)-\mu_{j}\right]\mathbf{p}_{j}\left(\tau\right)\text{, \ \ \ \ }\mathbf{p}_{j}\left(0\right)=\mathbf{v}_{j},\\
\mathbf{\dot{q}}_{j}^{\dagger}\left(\tau\right) & =\mathbf{q}_{j}^{\dagger}\left(\tau\right)\left[\mu_{j}-\mathcal{L}\left(\tau\right)\right]\text{, \ \ \ \ }\mathbf{q}_{j}^{\dagger}\left(0\right)=\mathbf{w}_{j}^{\dagger},
\end{align}
\end{subequations} and the orthogonality conditions $\mathbf{q}_{j}^{\dagger}(\tau)\mathbf{p}_{l}(\tau)=\delta_{jl}\;\forall\tau$,
as easily proven from their definition.

Let us now comment on the general properties of this eigensystem,
which we prove in detail in \cite{SupMat}. There always exists a
null eigenvalue, say $\mu_{0}=0$, with related (right) Floquet eigenvector
$\mathbf{p}_{0}(\tau)$. This property is a byproduct of the spontaneous
temporal symmetry breaking generated by the limit cycle (Goldstone
theorem). In the single-mode case, there is only one other eigenvalue,
which is given by $\mu_{1}=\int_{0}^{T}\frac{d\tau}{T}\text{tr}\{\mathcal{L}(\tau)\}$,
and has associated (left) Floquet eigenvector $\mathbf{q}_{1}(\tau)=(-\mathrm{i}\partial_{\tau}\bar{\beta},\mathrm{i}\partial_{\tau}\bar{\beta}^{*})^{T}\exp\left\{ \int_{0}^{\tau}d\tau'\text{tr}\{\mathcal{L}(\tau')\}-\mu_{1}\tau\right\} $.
This vector is the temporal counterpart of the linear or angular momentum
found in previous works which deal with spatial symmetries \cite{CNBthesis}.

Note that $\mathbf{p}_{0}(\tau)$ and $\mathbf{q}_{1}(\tau)$ are,
respectively, the tangent and normal vectors of the limit cycle's
trajectory, see Fig.~\ref{Fig1}(a). We haven't found explicit expressions
of the other Floquet eigenvectors in terms of the $\bar{\beta}(\tau)$,
but they can always be found numerically in an efficient fashion,
as we do for Fig.~\ref{Fig1}(a).

\textbf{Diffusion of the temporal pattern. }As a first physical consequence
of the properties above, we now show that $\theta$ is diffusing due
to quantum noise, and hence quantum fluctuations smear off the classical
periodic orbit.

In order to show this, we just need to apply $\mathbf{q}_{0}^{\dagger}(\tau)$
on (\ref{LinLan}), obtaining $\frac{d}{d\tau}(\mathbf{q}_{0}^{\dagger}\mathbf{b}+\theta)=\sqrt{\gamma}\mathbf{q}_{0}^{\dagger}(\tau)\mathbf{n}(\tau)$.
Note that by taking $\theta$ as a variable in (\ref{StochasticExpansion})
we introduced a redundancy in the number of variables, which is now
consistently removed by setting $\mathbf{q}_{0}^{\dagger}\mathbf{b}=0$
(in other words, introducing $\theta$ simply allowed us to track
and give physical meaning to this part of the quantum fluctuations).
The previous equation turns then into a diffusion equation for $\theta$,
leading to a variance 
\begin{equation}
\langle[\theta(\tau)-\theta(0)]^{2}\rangle=\gamma\int_{0}^{\tau}d\tau'\mathbf{q}_{0}^{\dagger}(\tau')\mathcal{N}(\tau')\mathbf{q}_{0}^{*}(\tau').\label{ThetaVariance}
\end{equation}
Note that the kernel is periodic, and therefore, the coarse-grained
dynamics of $\theta$ corresponds to a diffusion process, with a variance
increasing linearly with time, making $\theta$ fully undetermined
asymptotically as shown in Fig.~\ref{Fig1}(c).

\begin{figure}
\includegraphics[width=1\columnwidth]{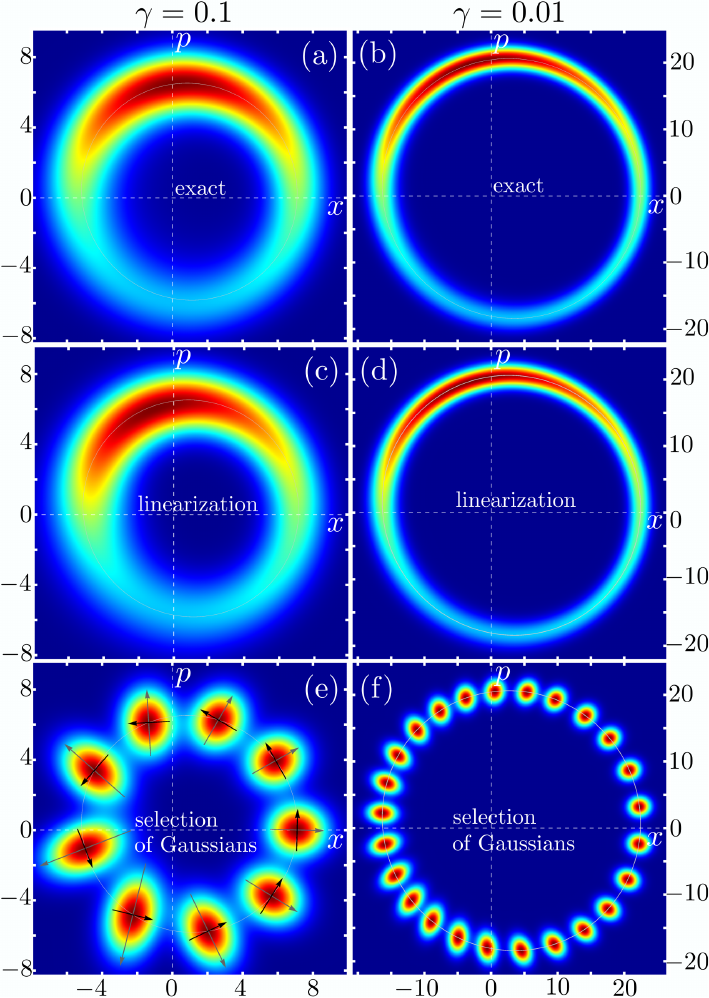}\caption{Steady-state Wigner functions of the driven VdP oscillator, Eq. (\ref{DPOmaster}),
for $\Delta=\sqrt{0.4}$, $F=\sqrt{0.1}$, and two values of $\gamma$,
0.1 and 0.01. In (a, b) we show the exact solutions, to be compared
with the linearized ones (c,d). In (e) and (f) we show a few of the
Gaussian states of Eq. (\ref{GaussianStates}) which we mix to form
the linearized approximation (\ref{FinalStateLinearized-1}). Note
how, as shown explicitly in (e), all the Gaussians carry vacuum fluctuations
along the direction defined by the $\mathbf{q}_{0}(\tau)$ Floquet
eigenvector (black arrows), with varying fluctuations along the $\mathbf{p}_{1}(\tau)$
direction (grey arrows).\label{Fig2}}
\end{figure}

\textbf{Steady state as a mixture of Gaussians.} The above considerations
imply that the steady state is formed by a balanced mixture of Gaussian
states, one for each value of $\theta$. As we prove below, the Wigner
functions of these Gaussian states \cite{CNBbook} are given by
\begin{equation}
W(\mathbf{r},\tau+\theta)=\frac{e^{-\frac{1}{2}[\mathbf{r}-\bar{\mathbf{d}}(\tau+\theta)]^{T}\bar{V}^{-1}(\tau+\theta)[\mathbf{r}-\bar{\mathbf{d}}(\tau+\theta)]}}{2\pi\sqrt{\det\{\bar{V}(\tau+\theta)\}}},\label{GaussianStates}
\end{equation}
where $\mathbf{r}=(x,p)^{T}$ is the coordinate vector in phase space,
and the mean vector and covariance matrix are given by \begin{subequations}\label{GaussianMoments}
\begin{align}
\bar{\mathbf{d}}(\tau) & =\mathcal{U}[\bar{\beta}(\tau),\bar{\beta}^{*}(\tau)]^{T}/\sqrt{\gamma},\label{GaussMean-1}\\
\bar{V}(\tau) & =\mathcal{I}+C(\tau)\mathcal{U}\mathbf{p}_{1}(\tau)\mathbf{p}_{1}^{T}(\tau)\mathcal{U}^{T}.\label{GaussCovar-1}
\end{align}
\end{subequations}$\mathcal{U}=\tiny{\left(\begin{array}{cc}
1 & 1\\
-\mathrm{i} & \mathrm{i}
\end{array}\right)}$ is the matrix that connects the complex representation of the bosonic
mode to its real representation in phase space, and
\begin{equation}
C(\tau)=\lim_{\tau\rightarrow\infty}\int_{0}^{\tau}d\tau'e^{2\mu_{1}(\tau-\tau')}\mathbf{q}_{1}^{\dagger}(\tau')\mathcal{N}(\tau')\mathbf{q}_{1}^{*}(\tau'),
\end{equation}
is a $T$-periodic function.

Let us now prove the expressions above. First, we introduce the quadrature
vector $\mathbf{R}=\mathcal{U}(\beta,\beta^{+})^{T}/\sqrt{\gamma}$.
Within the positive \emph{P} representation the elements of the long-time
mean vector $\bar{\mathbf{d}}$ and and covariance matrix $\bar{V}$\textbf{
}are found as $\bar{d}_{m}(\tau)=\lim_{\tau\rightarrow\infty}\langle R_{m}(\tau)\rangle$
and $\bar{V}_{mn}(\tau)=\delta_{mn}+\lim_{\tau\rightarrow\infty}\langle\delta R_{m}(\tau)\delta R_{n}(\tau)\rangle$,
where $\delta R_{m}=R_{m}-\langle R_{m}\rangle$ \cite{CNBbook}.
Next, note that the condition $\mathbf{q}_{0}^{\dagger}(\tau)\mathbf{b}(\tau)=0$
allows us to write the quantum fluctuations as $\mathbf{b}(\tau)=c_{1}(\tau)\mathbf{p}_{1}(\tau)$,
where we define the projection $c_{1}(\tau)=\mathbf{q}_{1}^{\dagger}(\tau)\mathbf{b}(\tau)$.
Using the expansion (\ref{StochasticExpansion}), we can then write
the quadrature vector as $\sqrt{\gamma}\mathbf{R}(\tau)=\mathcal{U}[\bar{\beta}(\tau),\bar{\beta}^{+}(\tau)]^{T}+c_{1}(\tau)\mathcal{U}\mathbf{p}_{1}(\tau)$,
whose stochastic properties are all then concentrated on $c_{1}(\tau)$.
On the other hand, applying $\mathbf{q}_{1}^{\dagger}(\tau)$ on (\ref{LinLan})
we find $\dot{c}_{1}=\mu_{1}c_{1}+\sqrt{\gamma}\mathbf{q}_{1}^{\dagger}(\tau)\mathbf{n}(\tau)$,
whose solution leads to the moments $\lim_{\tau\rightarrow\infty}\langle c_{1}(\tau)\rangle=0$
and $\lim_{\tau\rightarrow\infty}\langle c_{1}^{2}(\tau)\rangle=\gamma C(\tau)$,
which provide the mean vector and covariance matrix in (\ref{GaussianMoments}).

The steady state associated to the expansion (\ref{StochasticExpansion})
of the stochastic variables is then given by the balanced mixture
\begin{equation}
\bar{W}(\mathbf{r})=\int_{0}^{T}\frac{d\theta}{T}W(\mathbf{r};\tau+\theta)=\int_{0}^{T}\frac{d\theta}{T}W(\mathbf{r};\theta).\label{FinalStateLinearized-1}
\end{equation}
In Fig.~\ref{Fig2} we compare the Wigner function (\ref{FinalStateLinearized-1})
with the one obtained by exact simulation \cite{CNBnumericsNotes}
of the master equation (\ref{DPOmaster}). We find very good agreement
even for relatively large $\gamma$, where quantum fluctuations are
still quite relevant, as can be appreciated. 

This Wigner function has a very suggestive interpretation, see Fig.~\ref{Fig2}.
First, (\ref{GaussMean-1}) tells us that the Gaussian states are
centered along the points of the limit cycle's trajectory, as expected.
As for quantum fluctuations, note that the eigenvalues of the covariance
matrix $\bar{V}(\theta)$ are 1 and $\mathrm{det}\{\bar{V}(\theta)\}$,
which inform us about the variance along the principal axes of the
uncertainty ellipse. It is easy to check that the directions of these
principal axes follow the vectors $\mathcal{U}\mathbf{q}_{0}(\theta)$
and $\mathcal{U}\mathbf{p}_{1}(\theta)$ for the 1 and $\mathrm{det}\{\bar{V}(\theta)\}$
eigenvalues, respectively (see Fig.~\ref{Fig2}). Hence, the quadrature
of the Gaussian state which goes in the direction of $\mathbf{q}_{0}(\theta)$
(Goldstone mode) carries vacuum fluctuations, which one can trace
back to the condition $\mathbf{q}_{0}^{\dagger}(\theta)\mathbf{b}(\theta)=0$
that the method naturally demands. On the other hand, since in principle
all physical covariance matrices satisfy $\mathrm{det}\{\bar{V}\}\geq1$
(uncertainty principle) \cite{CNBbook}, this seems to suggest that
the quadrature going in the direction of $\mathbf{p}_{1}(\theta)$
carries fluctuations above the shot noise limit. While this is indeed
the case for the VdP oscillator studied here, our experience with
other nonlinear systems \cite{CNBthesis} tells us that we could find
$\mathrm{det}\{\bar{V}(\theta)\}<1$ (squeezing below shot noise)
without violating the uncertainty principle. This is because the two
quadratures of each Gaussian state are not conjugate variables, but
they are both conjugate to the diffusing variable $\theta$ \cite{CNBthesis},
which is completely undetermined in the steady state.

\textbf{Conclusions.} In this Letter we have introduced a linearization
method capable of dealing with quantum nonlinear systems in the regime
where they show spontaneous limit-cycle formation. The technique keeps
the simplicity of standard linearization around stationary solutions.
It requires finding the fundamental matrix of the Floquet method over
a period of the cycle by solving a linear initial value problem with
time-periodic coefficients. Only two equations are added with each
mode that is introduced in the problem, giving the method a linear
scaling with the size of the system that makes it suitable for complex
driven-dissipative many-body problems such as optomechanical arrays
\cite{Arrays1,Arrays2,Arrays3,Arrays4,ArraysWeiss}. Moreover, the
linearity of the equations should give efficient access also to dynamical
objects such as multi-time correlation functions, which are of crucial
relevance for experiments \cite{GardinerZoller,Carmichael1,CarmichaelBook2,GardinerZollerI}
and the emergent field of quantum synchronization \cite{Lorch16,Lee13,Walter14,Weiss17,Walter15,Mari13}. 
\begin{acknowledgments}
We thank Florian Marquardt for important suggestions and comments.
Our work also benefited from discussions with Alessandro Farace, Alejandro
González-Tudela, and Eugenio Roldán. This work was supported by the
ERC starting grant OPTOMECH, and by the Ministerio de Economa y Competitividad
of the Spanish Government and the European Union FEDER through project
FIS2014-60715-P.
\end{acknowledgments}

\begin{center}
\textbf{\newpage{}}
\par\end{center}
\begin{widetext}
\begin{center}
\textbf{Supplemental material}
\par\end{center}
This supplemental material is divided in three sections. In the first
one we derive the stochastic Langevin equations associated with the
master equation of the driven Van der Pol oscillator, and proceed
to their linearization. The second section is devoted to proving the
properties of the Floquet eigensystem that we introduced in the main
text. In the last section we provide a detailed overview of the phase
diagram of the Van der Pol oscillator in the classical limit.
\begin{center}
\textbf{\large{}I. Derivation of the stochastic equations}
\par\end{center}{\large \par}
The positive \textit{P} representation of a (single-mode) state $\hat{\rho}(\tau)$
is defined by \cite{Drummond80,CarmichaelBook2,GardinerZoller,GardinerZollerI}
\begin{equation}
\hat{\rho}(\tau)=\int_{\mathbb{C}^{2}}d^{4}\boldsymbol{\mathbf{\alpha}}P(\boldsymbol{\mathbf{\alpha}};\tau)\underset{\hat{\Lambda}(\boldsymbol{\mathbf{\alpha}})}{\underbrace{\frac{|\alpha\rangle\langle\alpha^{+\ast}|}{\langle\alpha^{+\ast}|\alpha\rangle}}},\label{StateFromP}
\end{equation}
with $\boldsymbol{\mathbf{\alpha}}=(\alpha,\alpha^{+})^{T}$. The
distribution $P(\boldsymbol{\mathbf{\alpha}};\tau)$ can always be
chosen as a well behaved positive distribution (see below), what is
accomplished at the expense of doubling the phase space of the oscillator,
since $\alpha$ and $\alpha^{+}$ are two independent complex variables.
Moments in normal order can be evaluated as \cite{Drummond80,CarmichaelBook2,GardinerZoller,GardinerZollerI}
\begin{equation}
\langle\hat{a}^{\dagger m}\hat{a}^{n}\rangle=\int_{\mathbb{C}^{2}}d^{4}\boldsymbol{\mathbf{\alpha}}P(\boldsymbol{\mathbf{\alpha}})\alpha^{+m}\alpha^{n}.\label{ExpectationsFromP}
\end{equation}
The master equation can be turned into a Fokker-Planck equation for
the distribution $P(\boldsymbol{\mathbf{\alpha}};\tau)$ as follows.
First, we introduce (\ref{StateFromP}) in the master equation, Eq.
(1) of the main text in our case, and use the properties
\begin{equation}
\hat{a}\hat{\Lambda}=\alpha\hat{\Lambda}\text{, \ \ }\hat{\Lambda}\hat{a}^{\dagger}=\alpha^{+}\hat{\Lambda}\text{, \ \ }\hat{\Lambda}\hat{a}=(\alpha+\partial_{\alpha^{+}})\hat{\Lambda}\text{, \ \ }\hat{a}^{\dagger}\hat{\Lambda}=(\alpha^{+}+\partial_{\alpha})\hat{\Lambda},
\end{equation}
leading to an equation of the form
\begin{equation}
\int_{\mathbb{C}^{2}}d^{4}\boldsymbol{\mathbf{\alpha}}\hat{\Lambda}(\boldsymbol{\mathbf{\alpha}})\partial_{\tau}P(\boldsymbol{\mathbf{\alpha}};\tau)=\int_{\mathbb{C}^{2}}d^{4}\boldsymbol{\mathbf{\alpha}}P(\boldsymbol{\mathbf{\alpha}};\tau)\left[\sum_{j=\alpha,\alpha^{+}}A_{j}(\boldsymbol{\mathbf{\alpha}})\partial_{j}+\frac{1}{2}\sum_{j,l=\alpha,\alpha^{+}}D_{jl}(\boldsymbol{\mathbf{\alpha}})\partial_{j}\partial_{l}\right]\hat{\Lambda}(\boldsymbol{\mathbf{\alpha}}).\label{PreFokkerPlanck}
\end{equation}
Here $A_{j}$ and $D_{jl}$ are the components of the drift vector
and the diffusion matrix, respectively, which in our case are found
to be\begin{subequations} 
\begin{align}
\mathbf{A} & =\left(\begin{array}{c}
(1+\mathrm{i}\Delta-\gamma\alpha^{+}\alpha)\alpha+F/\sqrt{\gamma}\\
(1-\mathrm{i}\Delta-\gamma\alpha^{+}\alpha)\alpha^{+}+F/\sqrt{\gamma}
\end{array}\right),\\
\mathcal{D} & =\left(\begin{array}{cc}
-\gamma\alpha^{2} & 2\\
2 & -\gamma\alpha^{+2}
\end{array}\right).
\end{align}
\end{subequations}Note that the analyticity of $\hat{\Lambda}(\boldsymbol{\alpha})$
gives us certain freedom to choose how the the complex derivatives
$\partial_{\alpha}$ and $\partial_{\alpha^{+}}$ act on it, what
can be used to always get a positive semidefinite diffusion matrix
\cite{CNBthesis,Drummond80}. Integrating by parts the right-hand
side of Eq. (\ref{PreFokkerPlanck}), and neglecting boundary terms
under the physical assumption that the distribution $P(\boldsymbol{\alpha};\tau)$
decays fast enough, we obtain the Fokker-Planck equation
\begin{equation}
\partial_{\tau}P(\boldsymbol{\mathbf{\alpha}};\tau)=\left[-\sum_{j=\alpha,\alpha^{+}}\partial_{j}A_{j}(\boldsymbol{\mathbf{\alpha}})+\frac{1}{2}\sum_{j,l=\alpha,\alpha^{+}}\partial_{j}\partial_{l}D_{jl}(\boldsymbol{\mathbf{\alpha}})\right]P(\boldsymbol{\mathbf{\alpha}};\tau).
\end{equation}

This equation is equivalent to the following stochastic \textit{\emph{Langevin}}\emph{
}equations \cite{Drummond80,CarmichaelBook2,GardinerZoller,GardinerZollerI,GardinerBook}
\begin{equation}
\boldsymbol{\dot{\alpha}}=\mathbf{A}+\mathcal{B}\boldsymbol{\eta}(\tau)\text{,}\label{StochasticLangevinAppendix}
\end{equation}
where $\mathcal{B}$ is a $2\times N$ matrix called the \textit{noise
matrix} which satisfies $\mathcal{BB}^{T}=\mathcal{D}$, and $\boldsymbol{\eta}(\tau)$
a vector whose $N$ components are independent real white Gaussian
noises ($N$ can be chosen at will, see below). Given the solution
$\boldsymbol{\alpha}[\tau;\boldsymbol{\eta}]$ as a functional of
the noises, the equivalence must be understood in a statistical sense
as
\begin{equation}
\langle\hat{a}^{\dagger m}\hat{a}^{n}\rangle=\int_{\mathbb{C}^{2}}d^{4}\boldsymbol{\mathbf{\alpha}}P(\boldsymbol{\mathbf{\alpha}};\tau)\alpha^{+m}\alpha^{n}=\langle\alpha^{+m}[\tau;\boldsymbol{\eta}]\alpha^{n}[\tau;\boldsymbol{\eta}]\rangle_{\text{stochastic}},\label{ExpectationsFromP2}
\end{equation}
that is, averaging over the distribution equals averaging over stochastic
realizations. In the following we remove the ``stochastic'' label
from the average, since the context will never allow confusing it
with quantum expectation values of operators.

As mentioned above, the ``internal'' dimension $N$ of the noise
matrix $\mathcal{B}$ is arbitrary in expression (\ref{StochasticLangevinAppendix}).
In general, it is possible to find a square noise matrix ($N=2$ in
our case), but sometimes it is simpler (or even more physical) to
work with $N>2$. In particular, in our case, we choose to work with
the noise matrix
\begin{equation}
\mathcal{B}=\left(\begin{array}{cccc}
\mathrm{i}\sqrt{\gamma}\alpha & 0 & 1 & \mathrm{i}\\
0 & -\mathrm{i}\sqrt{\gamma}\alpha^{+} & 1 & -\mathrm{i}
\end{array}\right),
\end{equation}
leading to the Langevin equations\begin{subequations} 
\begin{align}
\dot{\alpha} & =\frac{F}{\sqrt{\gamma}}+(1+\mathrm{i}\Delta-\gamma\alpha^{+}\alpha)\alpha+\mathrm{i}\sqrt{\gamma}\alpha\eta(\tau)+\sqrt{2}\xi(\tau),\\
\dot{\alpha}^{+} & =\frac{F}{\sqrt{\gamma}}+(1-\mathrm{i}\Delta-\gamma\alpha^{+}\alpha)\alpha^{+}-\mathrm{i}\sqrt{\gamma}\alpha^{+}\eta^{+}(\tau)+\sqrt{2}\xi^{\ast}(\tau),
\end{align}
\end{subequations}where $\eta(\tau)$, $\eta^{+}(\tau)$, and $\xi(\tau)$
are independent white Gaussian noises (real the first two, and complex
the last one), that is, they have zero average and 
\begin{equation}
\langle\eta(\tau)\eta(\tau^{\prime})\rangle=\langle\eta^{+}(\tau)\eta^{+}(\tau^{\prime})\rangle=\langle\xi(\tau)\xi^{\ast}(\tau^{\prime})\rangle=\delta(\tau-\tau^{\prime}),
\end{equation}
are their only nonzero two-time correlators.

It is finally interesting to rewrite the equations in terms of new
rescaled variables $\beta=\sqrt{\gamma}\alpha$ and $\beta^{+}=\sqrt{\gamma}\alpha^{+}$,
which read\begin{subequations} \label{LangevinEqsSup}
\begin{align}
\dot{\beta} & =F+(1+\mathrm{i}\Delta-\beta^{+}\beta)\beta+\sqrt{\gamma}[\sqrt{2}\xi(\tau)+\mathrm{i}\beta\eta(\tau)],\\
\dot{\beta}^{+} & =F+(1-\mathrm{i}\Delta-\beta^{+}\beta)\beta^{+}+\sqrt{\gamma}[\sqrt{2}\xi^{\ast}(\tau)-\mathrm{i}\beta^{+}\eta^{+}(\tau)].
\end{align}
\end{subequations}These are the equations that we provided in Eqs.
(2) of the main text. We took them as a starting point to present
the linearization technique, which we show in detail next.

The general linearization technique for dissipative systems affected
by spontaneous breaking of a continuous symmetry starts by applying
the symmetry transformation to the system, but with a parameter that
is allowed to vary in time \cite{CNBthesis}. In the present case,
the method finds the additional difficulty that the symmetry transformation
is a shift in time $\tau\rightarrow\tau+\theta$, and if the parameter
$\theta$ is to depend on time, the shift must be applied on it as
well. Technically, this makes it an infinitely-iterated function $\theta(\tau+\theta(\tau+\theta(...)))$,
which makes the derivation more elaborate than in previous systems
\cite{CNBthesis}. The (time-shifted) stochastic amplitudes are expanded
as the classical limit cycle plus some small quantum that can be assumed
to be small,
\begin{equation}
\beta(\tau+\theta)=\bar{\beta}(\tau+\theta)+b(\tau+\theta),\hspace{1em}\beta^{+}(\tau+\theta)=\bar{\beta}^{\ast}(\tau+\theta)+b^{+}(\tau+\theta),
\end{equation}
where we have omitted the time dependence of $\theta$ for ease of
notation.

When plugging this expressions into the stochastic Langevin equations
(\ref{LangevinEqsSup}), it is important to keep in mind that the
derivative of $\theta$ can be assumed small, since the method tells
us self-consistently that they are directly proportional to quantum
noise, see the paragraph before Eq. (8) in the main text. This means
that we can approximate
\begin{align}
\frac{d}{d\tau}\theta(\tau+\theta(\tau+\theta(...))) & =\partial_{\tau}\theta(\tau+\theta(\tau+\theta(...)))[1+[\partial_{\tau}\theta(\tau+\theta(\tau+\theta(...)))][1+[\partial_{\tau}\theta(\tau+\theta(\tau+\theta(...)))][1+...]]]\nonumber \\
 & \approx\partial_{\tau}\theta(\tau+\theta(\tau+\theta(...))),
\end{align}
 and therefore
\begin{align}
\frac{d}{d\tau}\beta(\tau+\theta(\tau+\theta(...))) & \approx[\partial_{\tau}\bar{\beta}(\tau+\theta(\tau+\theta(...)))+\partial_{\tau}b(\tau+\theta(\tau+\theta(...)))][1+\partial_{\tau}\theta(\tau+\theta(...))]\nonumber \\
 & \approx\partial_{\tau}\bar{\beta}(\tau+\theta(\tau+\theta(...)))\partial_{\tau}\theta(\tau+\theta(...))+\partial_{\tau}b(\tau+\theta(\tau+\theta(...))),
\end{align}
and similarly for $\beta^{+}$, where in the last line we have assumed
that the fluctuations $b$ and related derivatives are small. Introducing
these expansions into Eqs. (\ref{LangevinEqsSup}) evaluated at $\tau+\theta(\tau+\theta(...))$,
and keeping terms linear in noises and the small variables mentioned
above, we obtain the linearized Langevin equations introduced in Eq.
(4) of the main text, but time-shifted by $\theta(\tau+\theta(...))$.
The last step consists then in shifting the time arguments by $-\theta(\tau+\theta(...))$,
leading to the linearized equations as presented in the main text.

\bigskip{}

\bigskip{}

\bigskip{}

\begin{center}
\textbf{\large{}II. Floquet eigensystem}
\par\end{center}{\large \par}
In order to solve the linearized Langevin equations we applied the
Floquet method in the main text. In particular, we showed that all
the system properties are easy to derive from the Floquet eigensystem,
that is, from the eigenvalues $\{\mu_{j}\}_{j=0,1}$ and eigenvectors
satisfying\begin{subequations} \label{EigenTime-1}
\begin{align}
\mathbf{\dot{p}}_{j}\left(\tau\right) & =\left[\mathcal{L}\left(\tau\right)-\mu_{j}\right]\mathbf{p}_{j}\left(\tau\right)\text{, \ \ \ \ }\mathbf{p}_{j}\left(0\right)=\mathbf{v}_{j},\\
\mathbf{\dot{q}}_{j}^{\dagger}\left(\tau\right) & =\mathbf{q}_{j}^{\dagger}\left(\tau\right)\left[\mu_{j}-\mathcal{L}\left(\tau\right)\right]\text{, \ \ \ \ }\mathbf{q}_{j}^{\dagger}\left(0\right)=\mathbf{w}_{j}^{\dagger},
\end{align}
\end{subequations} and the orthogonality conditions $\mathbf{q}_{j}^{\dagger}(\tau)\mathbf{p}_{l}(\tau)=\delta_{jl}\;\forall\tau$.
In general, this has to be done numerically, especially since the
limit cycle itself does not admit analytic expressions except in special
situations. However, we mentioned in the main text a couple of analytic
properties of the eigensystem that we prove in this section.

Since these properties are not specific to the Van der Pol oscillator,
but general for any single-mode problem, let us consider a completely
general single-mode limit cycle $\bar{\beta}(\tau)$ satisfying the
equation
\begin{equation}
\frac{d\bar{\beta}}{d\tau}=A(\bar{\beta},\bar{\beta}^{*}),\label{GenLimitCycleEq}
\end{equation}
with associated linear stability matrix
\begin{equation}
\mathcal{L}(\tau)=\left(\begin{array}{cc}
\partial_{\bar{\beta}}A & \partial_{\bar{\beta}^{\ast}}A\\
\partial_{\bar{\beta}}A^{\ast} & \partial_{\bar{\beta}^{\ast}}A^{\ast}
\end{array}\right).\label{GenLinStaMat}
\end{equation}
 It is convenient to define the vector $\boldsymbol{\Pi}(\tau)=[\bar{\beta}(\tau),\bar{\beta}^{\ast}(\tau)]^{T}$.
Its first derivative satisfies the equation $\dot{\boldsymbol{\Pi}}=(A,A^{*})^{T}$,
leading to a second derivative obeying
\begin{equation}
\boldsymbol{\ddot{\Pi}}=\mathcal{L}(\tau)\boldsymbol{\dot{\Pi}},\label{PiEq}
\end{equation}
as is trivially proven from (\ref{GenLimitCycleEq}) and (\ref{GenLinStaMat}).

The first property we want to prove is the existence of one null eigenvalue,
say $\mu_{0}=0$, with an associated (right) Floquet eigenvector $\mathbf{p}_{0}=(\partial_{\tau}\bar{\beta},\partial_{\tau}\bar{\beta}^{\ast})^{T}$.
In order to prove it just note that according to (\ref{EigenTime-1})
$\mathbf{p}_{0}$ satisfies the equation $\mathbf{\dot{p}}_{0}\left(\tau\right)=\left[\mathcal{L}\left(\tau\right)-\mu_{0}\right]\mathbf{p}_{0}\left(\tau\right)$
by construction. On the other hand, since $\mathbf{p}_{0}=\boldsymbol{\dot{\Pi}}$,
we also have $\mathbf{\dot{p}}_{0}\boldsymbol{=}\mathcal{L}\left(\tau\right)\mathbf{p}_{0}$
by virtue of (\ref{PiEq}). Comparing these two expressions we obtain
$\mu_{0}=0$. This property finds its roots on the Goldstone theorem,
and indeed can be proven for an arbitrary number of modes by naturally
extending all the definitions. 

The other property that we provided in the main text was that the
other eigenvalue takes the value $\mu_{1}=\overline{\mathrm{tr}\{\mathcal{L}(\tau)\}}$,
where we define here $\overline{\mathrm{tr}\{\mathcal{L}(\tau)\}}=\int_{0}^{T}\frac{d\tau}{T}\text{tr}\{\mathcal{L}(\tau)\}$,
with associated (left) Floquet eigenvector 
\begin{equation}
\mathbf{q}_{1}(\tau)=\boldsymbol{\Pi}_{\mathrm{1}}\left(\tau\right)\exp\left\{ -\int_{0}^{\tau}d\tau^{\prime}\left[\mathrm{tr}\{\mathcal{L}(\tau)\}-\overline{\mathrm{tr}\{\mathcal{L}(\tau)\}}\right]\right\} ,\label{q1piLOF}
\end{equation}
with $\boldsymbol{\Pi}_{\mathrm{1}}\left(\tau\right)=[-\mathrm{i}\partial_{\tau}\bar{\beta}(\tau),\mathrm{i}\partial_{\tau}\bar{\beta}^{\ast}(\tau)]^{T}$.
The expression for the eigenvalue is readily proven by noticing that
for a general Floquet problem, the following property holds \cite{GrimshawBook,CodingtonBook}:
$\sum_{j}\mu_{j}=\overline{\mathrm{tr}\{\mathcal{L}(\tau)\}}$. Hence,
for a single-mode problem we obtain what we are looking for, since
there are only two eigenvalues and one of them is 0 as proven above. 

To prove that (\ref{q1piLOF}) is the corresponding eigenvector we
need to work a bit harder. We will proceed by making the ansatz $\mathbf{q}_{1}(\tau)=f(\tau)\boldsymbol{\Pi}_{\mathrm{1}}\left(\tau\right)$
for some real function $f(\tau)$, and proving that such a function
exists. It is convenient to remind ourselves of certain objects that
naturally appear in the Floquet method (see the main text for more
details). First, the fundamental matrix $\mathcal{R}$ which satisfies
the equation $\dot{\mathcal{R}}=\mathcal{L}(\tau)\mathcal{R}$. This
matrix defines a constant matrix $\mathcal{M}$ through $\exp\left(\mathcal{M}T\right)=\mathcal{R}(\tau)$,
and a periodic matrix $\mathcal{P}(\tau)=\mathcal{R}(\tau)\exp\left(-\mathcal{M}\tau\right)$.
The Floquet eigenvectors are then defined as $\mathbf{p}_{j}(\tau)=\mathcal{P}(\tau)\mathbf{v}_{j}$
and $\mathbf{q}_{j}^{\dagger}(\tau)=\mathbf{w}_{j}^{\dagger}\mathcal{P}^{-1}(\tau)$,
where $\{\mathbf{v}_{j},\mathbf{w}_{j};\mu_{j}\}_{j=0,1}$ is the
eigensystem of $\mathcal{M}$, composed of right and left orthogonal
eigenvectors satisfying $\mathcal{M}\mathbf{v}_{j}=\mu_{j}\mathbf{v}_{j}$
and $\mathbf{w}_{j}^{\dagger}\mathcal{M}=\mu_{j}\mathbf{w}_{j}^{\dagger}$.
With these definitions at hand, we start by noting that
\begin{equation}
\mathbf{q}_{1}^{\dagger}(\tau)=\mathbf{w}_{1}^{\mathbf{\dagger}}\mathcal{P}^{-1}(\tau)=\mathbf{w}_{1}^{\dagger}e^{\mathcal{M}\tau}\mathcal{R}^{-1}(\tau)=e^{\mu_{1}\tau}\mathbf{w}_{1}^{\dagger}\mathcal{R}^{-1}(\tau).
\end{equation}
The time derivative of this expression and our ansatz yields
\begin{equation}
\dot{f}\boldsymbol{\Pi}_{1}^{\dagger}+f\boldsymbol{\dot{\Pi}}_{1}^{\dagger}=\mu_{1}e^{\mu_{1}\tau}\mathbf{w}_{1}^{\dagger}\mathcal{R}^{-1}+e^{\mu_{1}\tau}\mathbf{w}_{1}^{\dagger}\frac{d}{d\tau}\mathcal{R}^{-1}=f\boldsymbol{\Pi}_{1}^{\dagger}\left[\mu_{1}+\mathcal{L}\right],
\end{equation}
where we have used that, by definition, the fundamental matrix satisfies
$\frac{d}{d\tau}\mathcal{R}^{-1}=\mathcal{R}^{-1}\mathcal{L}$. From
this expression, we see that $\boldsymbol{\dot{\Pi}}_{1}^{\dagger}$
can be written as
\begin{equation}
\boldsymbol{\dot{\Pi}}_{1}^{\dagger}=\boldsymbol{\Pi}_{1}^{\dagger}\left[\mu_{1}-\frac{\dot{f}}{f}-\mathcal{L}\right].\label{dPi1}
\end{equation}
The next step is obtaining $\boldsymbol{\dot{\Pi}}_{1}^{\dagger}$
following a different path. Let us define the matrix
\begin{equation}
\mathcal{J}=\left(\begin{array}{cc}
1 & 0\\
0 & -1
\end{array}\right),
\end{equation}
which allows us to write
\begin{equation}
\boldsymbol{\Pi}_{1}(\tau)=-\mathrm{i}\mathcal{J}\boldsymbol{\dot{\Pi}}(\tau)\Longrightarrow\boldsymbol{\dot{\Pi}}_{1}=-\mathrm{i}\mathcal{J}\boldsymbol{\ddot{\Pi}}=-\mathrm{i}\mathcal{JL}\boldsymbol{\dot{\Pi}=}\mathcal{JLJ}\boldsymbol{\Pi}_{1}.
\end{equation}
Next, we exploit the structure of any single-mode linear stability
matrix (\ref{GenLinStaMat}) to write
\begin{equation}
\mathcal{JLJ}=\mathrm{tr}\{\mathcal{L}\}\mathcal{I}-\mathcal{L}^{\dagger},
\end{equation}
which combined with the previous result leads to
\begin{equation}
\boldsymbol{\dot{\Pi}}_{1}^{\dagger}\boldsymbol{=\Pi}_{1}^{\dagger}\left[\mathrm{tr}\{\mathcal{L}(\tau)\}\mathcal{I}-\mathcal{L}(\tau)\right].\label{dPi2}
\end{equation}
Finally, comparing (\ref{dPi1}) and (\ref{dPi2}), and using the
expression that we found for $\mu_{1}$ we get
\begin{equation}
\frac{\dot{f}}{f}=\overline{\mathrm{tr}\{\mathcal{L}(\tau)\}}-\mathrm{tr}\{\mathcal{L}(\tau)\},
\end{equation}
which shows that there is a indeed a solution for the ansatz function,
\begin{equation}
f(\tau)=\exp\left\{ \int_{0}^{\tau}d\tau^{\prime}\left[\overline{\mathrm{tr}\{\mathcal{L}(\tau)\}}-\mathrm{tr}\{\mathcal{L}(\tau)\}\right]\right\} ,
\end{equation}
where we have taken $f(0)=1$ for definiteness. This completes the
proof of (\ref{q1piLOF}).

\bigskip{}

\bigskip{}

\begin{center}
\textbf{\large{}III. Phase diagram in the classical limit}
\par\end{center}{\large \par}
In this section we analyze in detail the properties of the driven
Van der Pol oscillator in the classical limit. In the main text, we
argued that the classical limit corresponds to $\gamma\rightarrow0$
in the stochastic Langevin equations. Let us show here, for completeness,
that these are indeed the equations that are obtained by assuming
the state of the system to be coherent at all times, $\hat{\rho}(\tau)=|\beta(\tau)/\sqrt{\gamma}\rangle\langle\beta(\tau)/\sqrt{\gamma}|$,
with a time-dependent amplitude $\beta(\tau)$ that will be our classical
variable (normalized to $\sqrt{\gamma}$ for convenience). In order
to find an evolution equation for $\beta$, we proceed as follows.
Using the master equation (1) of the main text, the expectation value
of any operator $\hat{A}$ is shown to evolve according to
\begin{equation}
\frac{d}{d\tau}\langle\hat{A}\rangle=\frac{F}{\sqrt{\gamma}}\langle[\hat{A},\hat{a}^{\dagger}]\rangle+\frac{F}{\sqrt{\gamma}}\langle[\hat{A},\hat{a}]\rangle+\mathrm{i}\Delta\langle[\hat{A},\hat{a}^{\dagger}\hat{a}]\rangle+\frac{\gamma}{2}\langle[\hat{a}^{\dagger2},\hat{A}]\hat{a}^{2}\rangle+\frac{\gamma}{2}\langle\hat{a}^{\dagger2}[\hat{A},\hat{a}^{2}]\rangle+\langle[\hat{a},\hat{A}]\hat{a}^{\dagger}\rangle+\langle\hat{a}[\hat{A},\hat{a}^{\dagger}]\rangle.
\end{equation}
Applied to the annihilation operator $\hat{a}$ and using the coherent
state ansatz, such that $\langle\hat{a}^{\dagger m}\hat{a}^{n}\rangle=\beta^{\ast m}\beta^{n}/\gamma^{(m+n)/2}$,
we obtain the equation of motion
\begin{equation}
\dot{\beta}=F+(\mathrm{i}\Delta+1-|\beta|^{2})\beta,\label{ClassEqMotion}
\end{equation}
which is precisely the one we introduced in the main text and coincides
with the stochastic Langevin equations in the $\gamma\rightarrow0$
limit. Note that there are only two parameters in this equation, which
fully characterize the phase diagram in this limit, as we show in
Fig.~\ref{Fig-PhaseDiagramVdP} (see below for the meaning of $I$).

\begin{figure}[t]
\begin{centering}
\includegraphics[width=0.8\textwidth]{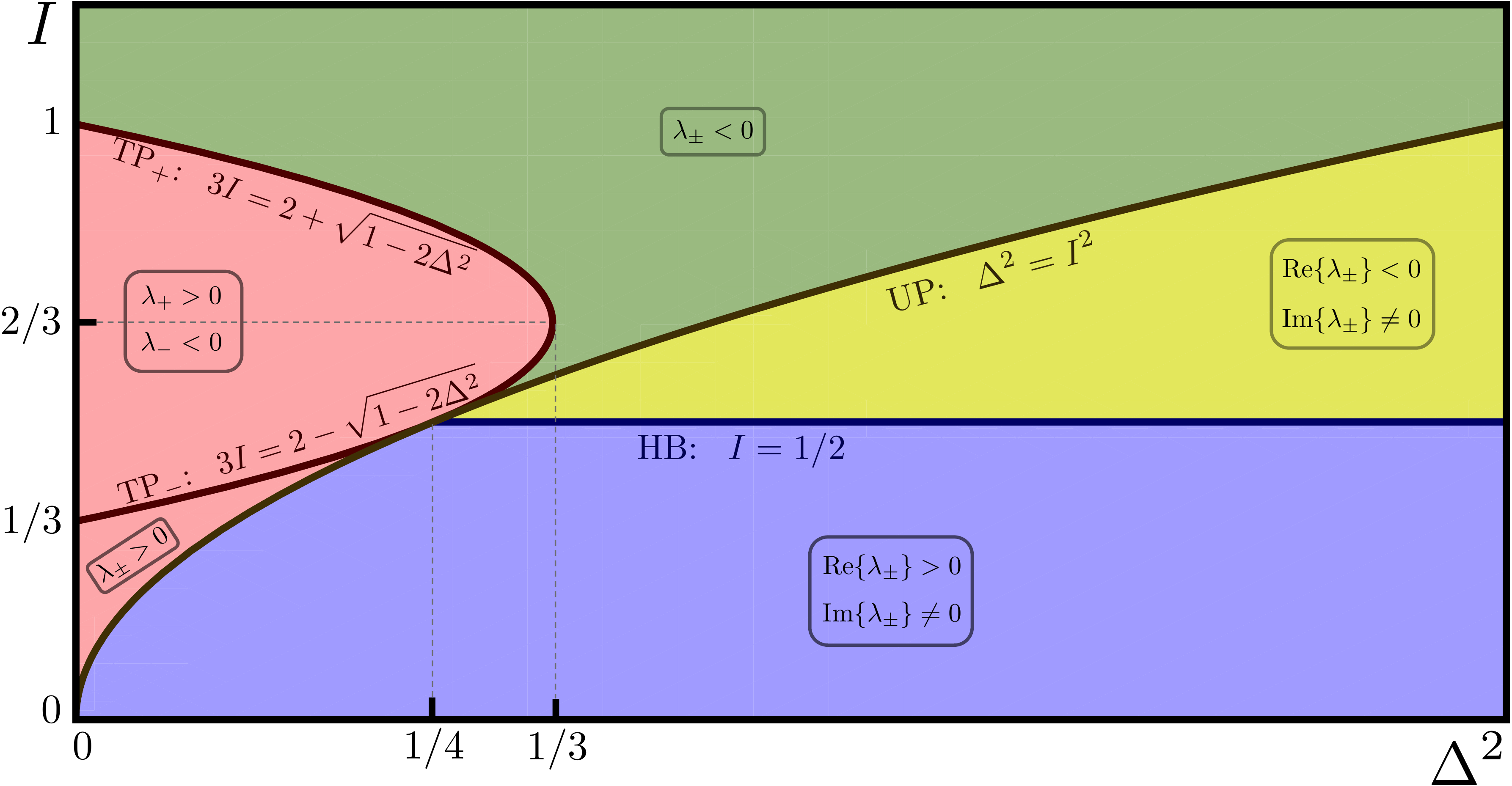} 
\par\end{centering}
\caption{Phase driagram of the driven Van der Pol oscillator in the classical
limit. The phase of the system is completely determined by two parameters,
namely the square of the detuning $\Delta^{2}$ and the oscillator
intensity $I$. The TP$_{\pm}$ curves correspond to the turning points
of the S-shaped response of the harmonic intensity to the injection
$F^{2}$ (see Fig. \ref{Fig-BifDiagramVdP}), which are both static
instabilities. The HB line corresponds to a Hopf bifurcation which
connects limit cycles with stationary solutions. The UP curve corresponds
to the points where the eigenvalues change from complex to real, which
for this system coincide with the points where phase oscillations
change from underdamped to overdamped. We see that this system offers
a wide variety of phases.}
\label{Fig-PhaseDiagramVdP}
\end{figure}

Depending on the parameters, the asymptotic (long-time term) solutions
of this equation may be time independent (stationary) or dependent
(limit cycles). In order to identify when these different regimes
happen, we first find the stationary asymptotic solutions $\bar{\beta}$
and study their stability. Let us write the amplitude as $\bar{\beta}=\sqrt{I}e^{\mathrm{i}\varphi}$,
with $I\in[0,\infty[$ and $\varphi\in[0,2\pi[$, which introduced
in the equation of motion (\ref{ClassEqMotion}) leads to the steady-state
equation $Fe^{-\mathrm{i}\varphi}=(I^{2}-1-\mathrm{i}\Delta)\sqrt{I}$,
or the equation for the \textit{\emph{oscillator}} \textit{intensity}
$I$
\begin{equation}
F^{2}=\left(\Delta^{2}+1\right)I-2I^{2}+I^{3},\label{IntensityF}
\end{equation}
from which the phase is recovered as $\varphi=\arg\{1-I^{2}-\mathrm{i}\Delta\}$.
This equation may possess one or several real and positive solutions,
depending on the parameters. In order to determine when each of these
possibilities occur, we simply determine the turning points $I=I_{\pm}$
of the S-shaped curve $I(F^{2})$ shown in Figs. \ref{Fig-BifDiagramVdP}.
These can be found as the extrema of $F^{2}(I)$, 
\begin{equation}
\left.\frac{\partial F^{2}}{\partial I}\right\vert _{I=I_{\pm}}=\left(\Delta^{2}+1\right)-4I_{\pm}+3I_{\pm}^{2}=0\text{ \ \ \ \ }\Longrightarrow\text{ \ \ \ \ }I_{\pm}=\frac{2\pm\sqrt{1-3\Delta^{2}}}{3}.
\end{equation}
Hence, we see that these points only exist when $\Delta^{2}<1/3$.
The values of the injection $F^{2}$ corresponding to these intensities
can be written as
\begin{equation}
F_{\pm}^{2}=\frac{2}{27}\left(2\pm\sqrt{1-3\Delta^{2}}\right)\left(1+3\Delta^{2}\mp\sqrt{1-3\Delta^{2}}\right).
\end{equation}
For injections between these two values, we then find three-valued
intensities.

\begin{figure}[t]
\begin{centering}
\includegraphics[width=0.9\textwidth]{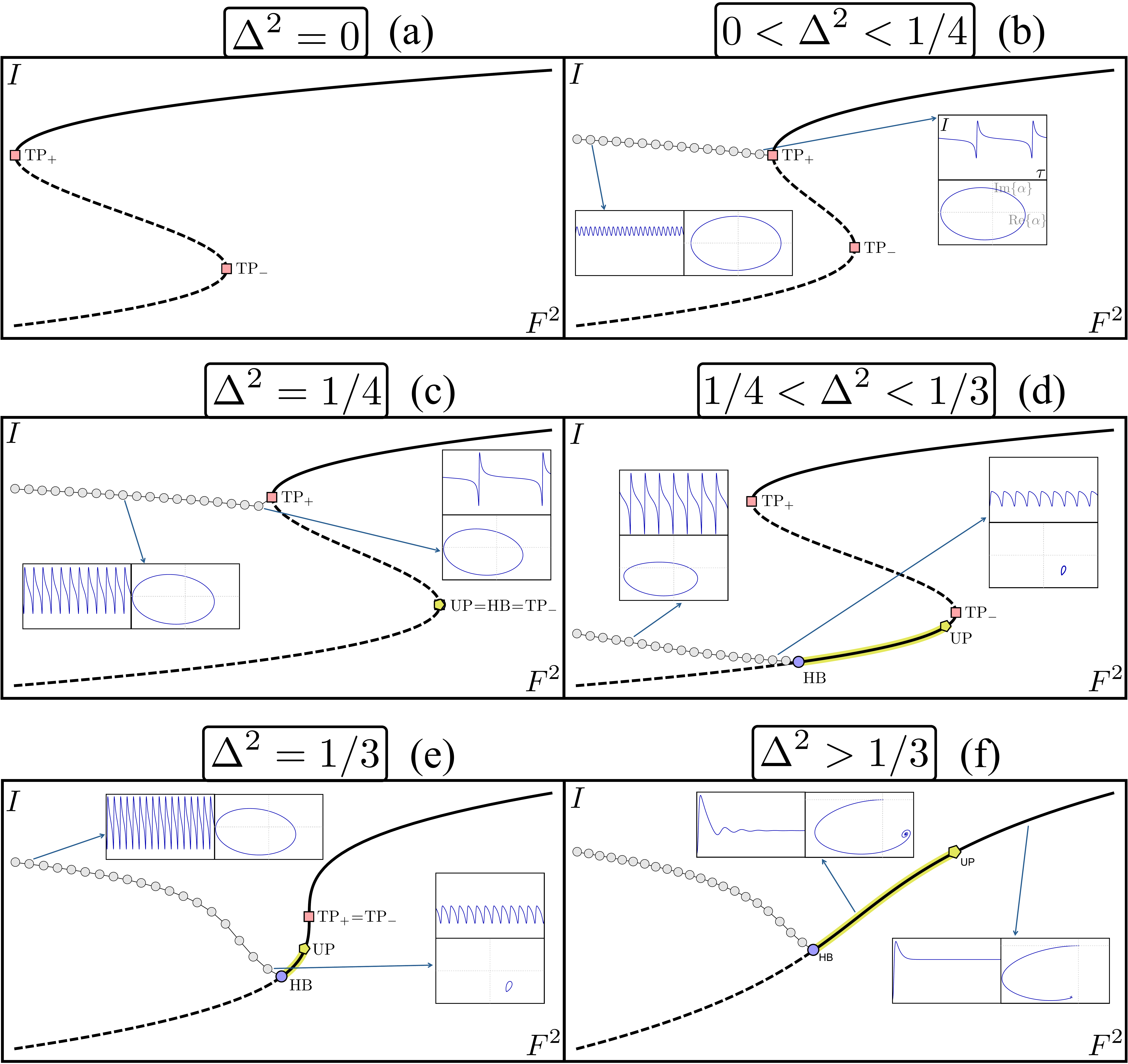} 
\par\end{centering}
\caption{Dynamical behaviour of the oscillator intensity as a function of the
injection $F^{2}$, in the $\Delta^{2}$-regions with distinct properties.
The solid lines provide the intensity $I$ of stable stationary amplitudes
with overdamped phase oscillations; the solid lines with yellow dashing
provide the same, but when phase oscillations are underdamped; the
dashed lines correspond to unstable stationary solutions; the grey
circles correspond to the mean intensity of the limit cycles which
we find numerically. The insets show the temporal dynamics of the
intensity $I$ (upper panel) as well as the trajectory of the limit
cycle in the phase space formed by the real and imaginary parts of
the amplitude $\beta$ (lower panel).}
\label{Fig-BifDiagramVdP}
\end{figure}

Let's now consider the stability of these solutions \cite{GrimshawBook,StrogatzBook}.
It is convenient to take the intensity $I$ as a parameter rather
than $F$, since the latter is uniquely determined from the former
through Eq. (\ref{IntensityF}), and not the other way around. In
order to analyze the stability of a stationary solution $\bar{\beta}$,
we write the amplitudes as $\beta(t)=\bar{\beta}+\delta\beta(t)$,
and consider terms up to linear order in the evolution equation (\ref{ClassEqMotion}).
Defining the vector $\boldsymbol{\beta}=\text{col}(\beta,\beta^{\ast})$,
this provides an evolution equation of the form $\delta\boldsymbol{\dot{\beta}}=\mathcal{L}\delta\boldsymbol{\beta}$,
where the \textit{linear stability matrix} reads
\begin{equation}
\mathcal{L}=\left(\begin{array}{cc}
1-2I+\mathrm{i}\Delta & -\bar{\beta}^{2}\\
-\bar{\beta}^{\ast2} & 1-2I-\mathrm{i}\Delta
\end{array}\right),\label{LSM-1}
\end{equation}
with characteristic polynomial
\begin{equation}
P(\lambda)=1-4I+3I^{2}+\Delta^{2}+2(2I-1)\lambda+\lambda^{2},
\end{equation}
and therefore eigenvalues
\begin{equation}
\lambda_{\pm}=1-2I\pm\sqrt{I^{2}-\Delta^{2}}.\label{Eigenvalues}
\end{equation}
Whenever the real part of at least one of these eigenvalues is positive,
the corresponding solution will be unstable. The points of the parameter
space where the real part of an eigenvalue is zero are known as \textit{instabilities}
or \textit{bifurcations} \cite{StrogatzBook}. It is customary to
start checking the simplest instabilities, those where the imaginary
part of the corresponding eigenvalue is zero as well, which we denote
by \emph{static instabilities}. In our case, it is readily shown that
the turning points $I=I_{\pm}$ are the only static instabilities
(see the curves marked as TP$_{\pm}$ in Fig. \ref{Fig-PhaseDiagramVdP}).
On the other hand, the instabilities can appear in eigenvalues with
nonzero imaginary parts, in which case we talk about \emph{Hopf bifurcations}
\cite{StrogatzBook}. In our case, imaginary parts in the eigenvalues
(\ref{Eigenvalues}) appear only when $I^{2}<\Delta^{2}$ (see the
curve marked as UP\footnote{UP stands for ``underdamped phase\textquotedblright{} oscillations,
a name that will get meaning later, when we show that the $I^{2}<\Delta^{2}$
region corresponds precisely to this regime of motion.} in Fig. \ref{Fig-PhaseDiagramVdP}). We then find a Hopf bifurcation
at $I=1/2$ (see the curve marked as HB in Fig. \ref{Fig-PhaseDiagramVdP}).
A careful analysis of the signs of the real part of the eigenvalues
in between these instability curves leads to the phase diagram shown
in Fig. \ref{Fig-PhaseDiagramVdP}. Note that we are able to draw
such a simple (but rich) phase diagram because we are dealing with
a single harmonic mode whose eigenvalues depend only on two parameters
($I$ and $\Delta^{2}$) and have simple analytic expressions.

It is also interesting to understand the behaviour of the oscillator's
amplitude as a function of the injection $F^{2}$ in the different
regions of the phase diagram, in particular for different values of
$\Delta^{2}$. This is what we can see in Fig. \ref{Fig-BifDiagramVdP}:

$\cdot$ It can be appreciated that for $\ \Delta=0$ only the upper
branch is stable, and increases monotonically with the injection from
$I=1$ at $F=0$ (Fig. \ref{Fig-BifDiagramVdP}a). Hence, the oscillator's
amplitude has a unique stationary response for all injections, since
the drive is in resonance with the oscillator's natural frequency.

$\cdot$ When $0<\Delta^{2}<1/4$ (Fig. \ref{Fig-BifDiagramVdP}b)
the upper turning point departs from $F=0$. The asymptotic solution
corresponds then to a limit cycle for $0<F<F_{+}$ and to a stationary
solution in the upper branch for $F>F_{+}$. The physical interpretation
is clear: as soon as the drive is detuned, synchronization of the
oscillator's oscillations to the drive requires a minimum value of
the injection in order to work. Note that for $F=0$, the limit cycles
are of the trivial form $\alpha(\tau)=e^{\mathrm{i}\Delta\tau+\mathrm{i}\varphi}/\gamma$,
which simply means that the amplitude oscillates at the natural frequency
of the oscillator.

$\cdot$ At $\Delta^{2}=1/4$ (Fig. \ref{Fig-BifDiagramVdP}c), both
the Hopf bifurcation HB and the UP point appear precisely at the lower
turning point. If the detuning is made larger, specifically $1/4<\Delta^{2}<1/3$
(Fig. \ref{Fig-BifDiagramVdP}d), both the Hopf bifurcation and the
UP point are located somewhere along the lower branch of the S-shaped
curve, the former always below the latter. The portion of the lower
branch in between HB and TP$_{-}$ becomes stable, with underdamped
phase oscillations in between HB and UP. We observe that in this regime
there is coexistence between the stationary solutions of the upper
branch, and either limit cycles connected to the HB from $F=0$ or
stationary solutions in the lower branch.

$\cdot$ For $\Delta^{2}=1/3$ the turning points coalesce (Fig. \ref{Fig-BifDiagramVdP}e),
and therefore for $\Delta^{2}\geq1/3$ the steady-state curve is no
longer S-shaped, but increases monotonically with the injection as
shown in Fig. \ref{Fig-BifDiagramVdP}f. We then identify a unique
behaviour of the amplitude for each value of the injection which can
correspond to limit cycles, underdamped phase oscillations, or overdamped
phase oscillations. 

It is interesting to note the different ways in which the limit cycles
converge to the stationary solutions in the different regimes, which
is what the insets allow us to discuss. In particular, when the limit
cycles connect with a static instability (such as in Figs. \ref{Fig-BifDiagramVdP}b
and c, where they connect with the upper turning point), their periodic
pattern has a longer stationary plateau the closer we get to the instability,
eventually reaching an infinite duration. On the other hand, when
the limit cycles connect with a Hopf instability (such as in Figs.
\ref{Fig-BifDiagramVdP}d, e, and f), their oscillation frequency
becomes closer to $\sqrt{\Delta^{2}-I^{2}}$ the closer they are to
the instability, while at the same time their oscillation amplitude
becomes smaller and smaller, eventually reaching zero.

Hence, we see that the VdP oscillator has a rich dynamical behaviour
in the classical limit.

One final thing left to prove is that, as mentioned above above, the
region where the eigenvalues are complex ($I^{2}<\Delta^{2}$) coincides
with the region of underdamped phase oscillations. In order to show
this, let us write the oscillator's amplitude in terms of intensity
and phase fluctuations around the steady state $\bar{\beta}$ as 
\begin{equation}
\beta(t)=\sqrt{I+\delta I(t)}e^{\mathrm{i}[\varphi+\delta\varphi(t)]}\underset{\delta\varphi,\delta I/I\ll1}{\approx}\bar{\beta}\left[1+\mathrm{i}\delta\varphi(t)+\frac{\delta I(t)}{2I}\right],
\end{equation}
so that the amplitude fluctuations can be written as $\delta\beta=\bar{\beta}(\mathrm{i}\delta\varphi+\delta I/2I)$.
Using now the form of the linear stability matrix (\ref{LSM-1}),
we then get from the real and imaginary parts of the linear system
for the fluctuations the following phase and amplitude equations\begin{subequations}
\begin{align}
\delta\dot{\varphi} & =(1-2I-I\cos\varphi)\delta\varphi+(I\sin\varphi+\Delta)\frac{\delta I}{2I},\\
\frac{\delta\dot{I}}{2I} & =(1-2I+I\cos\varphi)\frac{\delta I}{2I}+(I\sin\varphi-\Delta)\delta\varphi.
\end{align}
\end{subequations}These first order differential equations can be
easily recasted as the following second order differential equation
for the phase fluctuations 
\begin{equation}
\delta\ddot{\varphi}+\underset{\Gamma}{\underbrace{2(2I-1)}}\delta\dot{\varphi}+\underset{\Omega^{2}}{\underbrace{[\Delta^{2}+(2I-1)^{2}-I^{2}]}}\delta\varphi=0,
\end{equation}
which is the equation of a damped harmonic oscillator. Hence, the
condition for underdamped phase oscillations is $\Gamma^{2}<4\Omega^{2}$,
leading to $I^{2}<\Delta^{2}$, just as we wanted to prove.

\newpage{}
\end{widetext}

\end{document}